\documentclass{article}
\usepackage[T1]{fontenc} 
\usepackage[utf8]{inputenc} 
\usepackage{ismir,amsmath,cite,url}
\usepackage{graphicx}
\usepackage{xcolor}
\usepackage{array} 
\usepackage{subcaption} 
\usepackage{tabularx} 
\usepackage{multirow}
\usepackage{makecell}
\usepackage{stfloats}

\title{Towards Explainable Music Emotion Recognition:\\
	The Route via Mid-level Features}

\multauthor
{Shreyan Chowdhury \hspace{0.5cm} Andreu Vall \hspace{0.5cm} Verena Haunschmid \hspace{0.5cm} Gerhard Widmer} { 
Institute of Computational Perception, Johannes Kepler University Linz, Austria\\
{\tt\small {firstname.lastname}@jku.at}
}
\def\authorname{Shreyan Chowdhury, Andreu Vall, Verena Haunschmid, Gerhard Widmer}

\begin{document}

\maketitle
\begin{abstract}

Emotional aspects play an important part in our interaction with music. However, modelling these aspects in MIR systems have been notoriously challenging since emotion is an inherently abstract and subjective experience, thus making it difficult to quantify or predict in the first place, and to make sense of the predictions in the next. In an attempt to create a model that can give a musically meaningful and intuitive explanation for its predictions, we propose a VGG-style deep neural network that learns to predict emotional characteristics of a musical piece together with (and based on) human-interpretable, mid-level perceptual features. We compare this to predicting emotion directly with an identical network that does not take into account the mid-level features and observe that the loss in predictive performance of going through the mid-level features is surprisingly low, on average. The design of our network allows us to visualize the effects of perceptual features on individual emotion predictions, and we argue that the small loss in performance in going through the mid-level features is justified by the gain in explainability of the predictions.
\end{abstract}

\section{Introduction}
\label{sec:introduction}

Emotions -- portrayed, perceived, or induced -- are an important aspect
of music. MIR systems can benefit from leveraging this aspect because of its direct impact on human perception of music, but doing so has been challenging due to the inherently abstract and subjective quality of this feature. Moreover, it is difficult to interpret emotional predictions in terms of musical content.
In our quest for computer systems
that can give musically or perceptually meaningful justifications
for their predictions \cite{widmer2017getting}, we turn to the notion
of \textit{`mid-level perceptual features'} as recently described and advocated
by several researchers \cite{friberg2014using, aljanaki2018}. These are musical qualities
(such as rhythmic complexity, or perceived major/minor harmonic character)
that are supposed to be musically meaningful and intuitively recognizable by
most listeners, without requiring music-theoretic knowledge.
It has been shown previously that there is
considerable consistency in human perception of these features; that they
can be predicted relatively well from audio recordings; and that they also
relate to the perceived emotional qualities of the music \cite{aljanaki2018}.

That is the motivation for the work to be reported here. Our goal is to
use mid-level features as a basis for providing explanations of (and thus
get further insights into, or handles on) a model's emotion predictions,
by training it to recognize mid-level qualities from audio, and predict
emotion ratings from the mid-level predictions. Further, we wish to quantify the
cost -- in terms of loss of predictive performance -- incurred by this detour.
We will call this the `cost of explainability'.

\begin{figure}
	\centering
	\includegraphics[clip, trim=0.5cm 0.5cm 0cm 2cm, width=1.00\columnwidth]{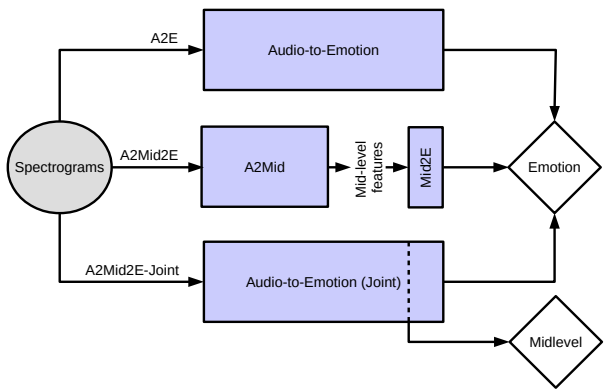}
	\caption{Three different architectures are compared for predicting emotion from audio.
	}
	\label{fig:models}
\end{figure}

Focusing our study on a specific benchmark dataset labelled with both perceived
emotional qualities and mid-level perceptual features, we first establish our basic VGG-style model architecture~\cite{simonyan2015}, showing that it can learn the two individual prediction tasks
(mid-level from audio, emotion from mid-level) from appropriate ground-truth data,
with accuracies that are at least on par with previously published models.
We then present a network with nearly identical architecture that learns to predict emotional
characteristics of a piece by explicitly going through a mid-level
feature prediction layer. 
We compare this to predicting emotion directly, using an identical
network with the exception of the mid-level layer, and find that the cost of going
through the mid-level features is surprisingly low, on average. Finally, we show that
by training the network to learn to predict mid-level and emotions
jointly, the results can be further improved. A graphical overview of the general
scenario is shown in Figure \ref{fig:models}.

The fact that in our model, emotions are predicted from the mid-level by a
single fully-connected layer, allows us to measure the effects of each of
the features on each emotion prediction, providing the basis for
interpretability and simple explanations. There are a number of application
scenarios in which we believe this could be useful; some of these will be briefly
discussed in the final section.

The remainder of this paper is structured as follows:
Section \ref{sec:relatedwork} briefly discusses related work on which this
research is based. In Section~\ref{sec:datasets} the datasets that provide us with emotion and mid-level annotations are described. The three different approaches to modelling emotion are summarized in Section~\ref{sec:models}. Experimental results and a demonstration of interpretability are given in Sections~\ref{sec:experiments} and ~\ref{sec:iml}. We discuss our findings and conclude in Section~\ref{sec:discussion}.

\section{Related Work}\label{sec:relatedwork}

In the MIR field, audio-based music emotion recognition (MER) has traditionally been done by extracting selected features from the audio and predicting emotion based on subsequent processing of these features~\cite{kim2010music}. Methods such as linear regression, regression trees, support vector regression, and variants have been used for prediction as mentioned in the systematic evaluation study by Huq et al~\cite{huq2010automated}. Techniques using regression-like algorithms have generally focused on predicting arousal and valence as per the well-known Russell's \textit{circumplex model} of emotion~\cite{russell1980circumplex}. Deep learning methods have also been employed for predicting arousal and valence, for example~\cite{xu2015multi}, that investigated BLSTM-RNNs in tandem with other methods, and \cite{coutinho2015automatically}, that used LSTM-RNNs. Others such as \cite{panda2013multi} and \cite{lin2016music} use support vector classification to predict the emotion class.
Aljanaki et al.~\cite{aljanaki2017} provide a summary of entries to the MediaEval emotion characterization challenge and quote results for arousal and valence prediction.

Deep neural networks are preferable for many tasks due to their high performance but can be considered black boxes due to their non-linear and nested structure. While in some fields such as healthcare or criminal justice the use of predictive analytics can have life-affecting consequences~\cite{rudin2018}, the decisions of MIR models are generally not as severe. Nevertheless, also in MIR it would be desirable to be able to obtain explanations for the decisions of a music recommendation or search system, for various reasons (see also Section \ref{sec:discussion}). Many current methods for obtaining insights into deep network-based audio classification systems do not explain the predictions in a human understandable way but rather design special filters that can be visualized~\cite{ravanelli2018sincnet}, or analyze neuron activations~\cite{krug2018irasl}. To the best of our knowledge, \cite{zhang2016} is the only attempt to build an interpretable model for MER. They performed the task of feature extraction and selection and built models from different model classes on top of them. The only interpretation offered is the reporting of coefficients from their logistic regression models,
without further explanation.

\section{Mid-level Perceptual Features}

The notion of \textit{(`mid-level') perceptual features} for characterizing music recordings has been put forward by several authors, as an alternative to purely sound-based or statistical low-level features (e.g., MFCCs, ZCR, spectral centroid) or more abstract music-theoretic concepts (e.g., meter, harmony). The idea is that they should represent musical characteristics that are easily perceived and recognized by most listeners, without any music-theoretical training. 
Research on such features has quite a history in the fields of music cognition and psychology (see \cite{friberg2014using} for a compact discussion).
Such features are attractive for our purposes because they could provide the basis for intuitive explanations of a MIR system's decisions, relating as they do to the musical experience of most listeners.

Various sets of such perceptual features have been proposed in the literature. For instance, Friberg et al.'s set \cite{friberg2014using} contains such concepts as speed, rhythmic clarity/complexity, articulation, dynamics, modality, overall pitch hight, etc.
In our study, we will be using the seven mid-level features defined by Aljanaki \& Soleymani \cite{aljanaki2018}, because they come with an openly available set of annotated audios (see below).
We recapitulate the features and their definitions in Table \ref{tbl:aljanakifeatures}, for convenience.

\begin{table*}
    \centering
	\begin{tabular}{l|c}
      \hline
      Perceptual Feature & Question asked to human raters \\
      \hline \hline
      Melodiousness & To which excerpt do you feel like singing along? \\
      \hline
      Articulation & Which has more sounds with staccato articulation? \\
      \hline
      Rhythmic Stability & \makecell{Imagine marching along with the music.\\Which is easier to march along with?} \\
      \hline
      Rhythmic Complexity & \makecell{Is it difficult to repeat by tapping?\\
Is it difficult to find the meter?\\
Does the rhythm have many layers?} \\
      \hline
      Dissonance & \makecell{Which excerpt has noisier timbre?\\
Has more dissonant intervals (tritones, seconds, etc.)?} \\
      \hline
      Tonal Stability & \makecell{Where is it easier to determine the tonic and key?\\
In which excerpt are there more modulations?} \\
      \hline
      Modality (`Minorness') & \makecell{Imagine accompanying this song with chords.\\
Which song would have more minor chords?} \\
      \hline \hline
	\end{tabular}
	\caption{Perceptual mid-level features as defined in \cite{aljanaki2018}, along with questions that were provided to
	human raters to help them interpret the concepts. (The ratings were collected in a pairwise comparison scenario.)
	In the following, we will refer to the last one (\textit{Modality}) as \textit{`Minorness'}, to make the
	core of the concept clearer.
}
	\label{tbl:aljanakifeatures}
\end{table*}

\section{Datasets}\label{sec:datasets}


For our experiments, we need music recordings annotated both with mid-level perceptual features, and with human ratings along some well-defined emotion categories. Our starting point is Aljanaki \& Soleymani's \emph{Mid-level Perceptual Features} dataset \cite{aljanaki2018}, which provides mid-level feature annotations. For the actual emotion prediction experiments, we then use the \textit{Soundtracks} dataset, which is contained in the Aljanaki collection as a subset, and comes with numeric emotion ratings along 8 dimensions.




\subsection{Mid-level Perceptual Features Dataset}
\label{sec:ds_midlevel}

The \textit{Mid-level Perceptual Features Dataset} \cite{aljanaki2018} consists of 5000 song snippets of around 15 seconds each annotated according to the seven mid-level descriptors listed in Table \ref{tbl:aljanakifeatures}.
The annotators were required to have some musical education and were selected based on passing a musical test. The ratings range from 1 to 10 and were scaled by a factor of 0.1 before being used for our experiments.

\subsection{Emotion Ratings: The Soundtracks Dataset}
\label{sec:ds_sountracks}

The \textit{Soundtracks (Stimulus Set 1)} dataset, published by Eerola and Vuoskoski~\cite{eerola2011}, consists of 360 excerpts from 110 movie soundtracks. The excerpts come with expert ratings for five categories following the discrete emotion model (happy, sad, tender, fearful, angry) and three categories following the dimensional model (valence, energy, tension). This makes it a suitable dataset for musically conveyed emotions~\cite{eerola2011}. The ratings in the dataset range from 1 to 7.83 and were scaled by a factor of 0.1 before being used for our experiments.
As stated above, all the songs in this set are also contained in the Mid-level Features Dataset, so that both kinds of ground truth are available.

\section{Audio-to-Emotion Models}\label{sec:models}

\begin{figure}
	\centering
	\includegraphics[width=1.00\columnwidth]{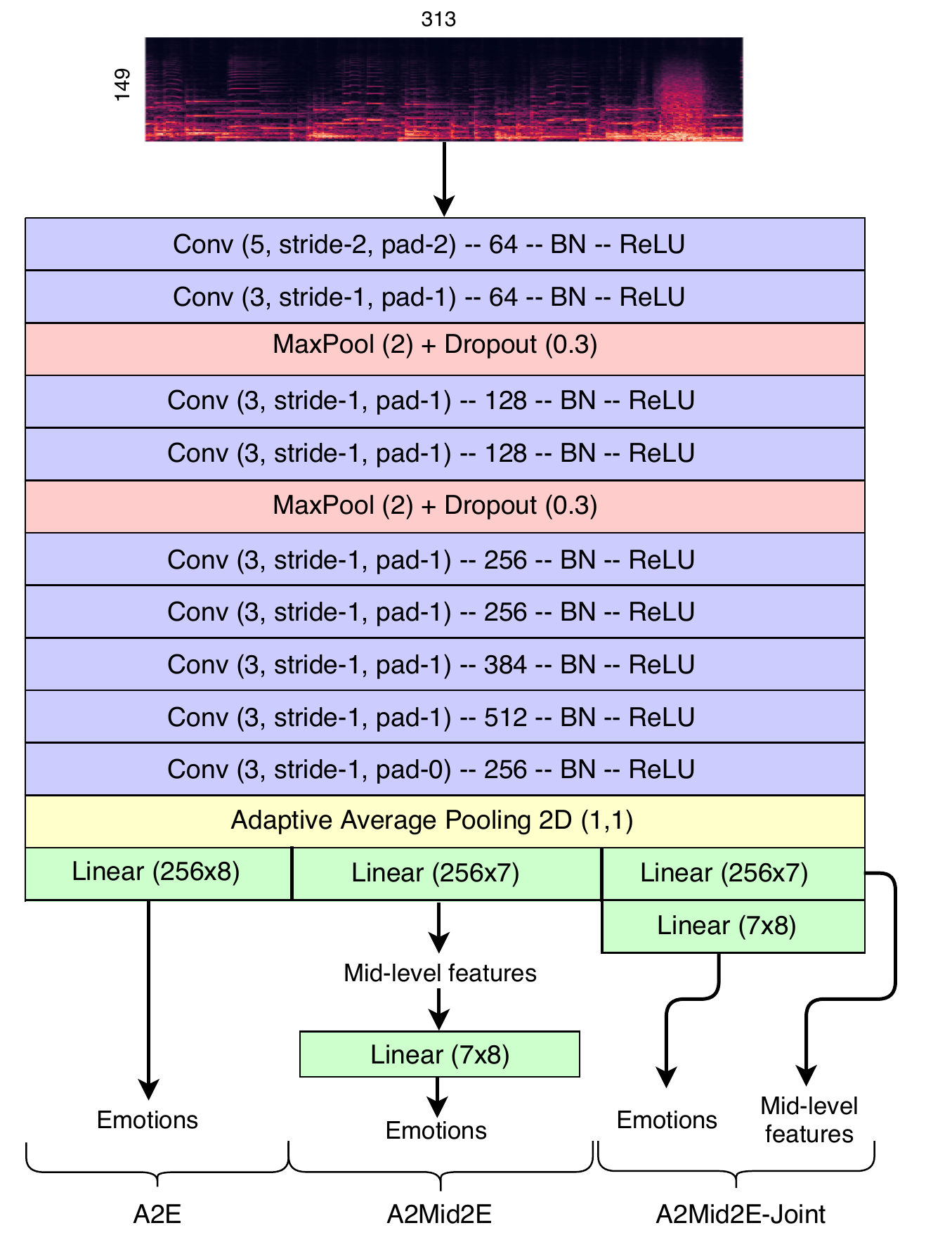}
	\caption{The same architecture from the first layer up to the `Adaptive Average Pooling 2D' layer is shared by all networks.}
	\label{fig:arch}
\end{figure}

In the following, we describe three different approaches to modeling emotion from audio, all based on VGG-style convolutional neural networks (CNNs). The architectures are summarized in Figure~\ref{fig:arch}. For all models, we use an Adam optimizer with a learning rate of 0.0005 and a batch size of 8, and employ early stopping with a patience of 50 epochs to prevent overfitting.

In terms of preprocessing, the audio samples are first converted into 149-point spectrograms calculated on randomly selected 10-second sections of the original snippets. The audio is resampled at 22.05 kHz, with a frame size of 2048 samples and a frame rate of 31.25 frames per second, and amplitude-normalized before computing the logarithmic-scaled spectrogram. This results in input vectors of size 313$\times$149.  These spectrograms are used as inputs for the following model architectures.

\subsection{A2E Scheme} 
\label{sec:a2emodel}

The first model, which we term ``A2E", is the most straightforward one. The spectrograms are fed into a VGG-style CNN to directly predict emotion values from audio. This is the leftmost path in Figure~\ref{fig:arch}. This model is not interpretable due to its black box architecture and is used as a baseline and for computing the \textit{cost of explainability} when comparing to more interpretable but possibly worse performing models.

\subsection{A2Mid2E Scheme} 
\label{sec:a2elinear}

In order to obtain a more interpretable model, an intermediate step is introduced. First, a VGG-style network is used to predict mid-level features from audio. This model is trained on the mid-level features dataset described in Section~\ref{sec:ds_midlevel} above. Next, a linear regression model is trained to predict the 8 emotion ratings in the Soundtracks dataset from the 7 mid-level feature values that we get as an output from the mid-level predictor network. This corresponds to a fully connected layer with 7 input units and 8 outputs and linear (identity) activation function -- see the middle path in Figure~\ref{fig:arch}. We call this scheme ``A2Mid2E". A linear model is chosen because its weights can easily be interpreted to understand the importance of each mid-level feature in predicting the emotion ratings.

\subsection{A2Mid2E-Joint Scheme} 
\label{sec:a2ejoint}

In an attempt to replace the step-wise training of two separate models with a single model that, ideally, could learn an internal representation useful for both prediction tasks, while keeping the interpretability of the linear weights, we propose a third architechture, called ``A2Mid2E-Joint" (rightmost path in Figure~\ref{fig:arch}). This network learns to predict mid-level features and emotion ratings jointly, but still predicts the emotions directly from the mid-level via a linear layer. This is achieved by the second last layer having exactly the same number of units as there are mid-level features (7), followed by a linear output layer with 8 outputs. From this network, we extract two outputs -- one from the second last layer ("mid-level layer"), and one from the last layer ("emotion layer"). We compute losses for both the outputs and optimize the combined loss (summation of both the losses).

\section{Experiments}
\label{sec:experiments}

\begin{table*}[ht]
    \centering
	\begin{tabular}{l|llllllll|l}
	    \hline
		~                       & Valence & Energy & Tension & Anger & Fear  & Happy & Sad   & Tender & Avg.     \\ \hline \hline
		Mid2E (Aljanaki)         & 0.88    & 0.79   & 0.84    & 0.65  & 0.82  & 0.81  & 0.73  & 0.72   & 0.78  \\ 
		Mid2E (Ours)             & 0.88   & 0.80  & 0.84   & 0.65 & 0.82 & 0.81 & 0.74 & 0.73  & 0.79 \\ \hline
		A2E                     & 0.81   & 0.79  & 0.84   & 0.82 & 0.81 & 0.66 & 0.60 & 0.75  & 0.76 \\
		A2Mid2E           & 0.79   & 0.74  & 0.78   & 0.72 & 0.77 & 0.64 & 0.58 & 0.67  & 0.71 \\
		A2Mid2E-Joint          & 0.82   & 0.78  & 0.82   & 0.76 & 0.79 & 0.65 & 0.64 & 0.72  & 0.75 \\ \hline \hline
		CoE$_{\text{A2Mid2E}}$ & 0.02   & 0.05 & 0.06   & 0.10 & 0.03 & 0.02 & 0.02 & 0.08  & 0.05 \\
		CoE$_{\text{A2Mid2E-Joint}}$  & -0.02   & 0.01  & 0.02   & 0.06 & 0.02 & 0.01 & -0.04 & 0.03  &0.01 \\ \hline
	\end{tabular}
	\caption{Summaries of the different model performances on predicting emotion. The last two rows show the ``cost of explainability'' (CoE), as the difference between our baseline (A2E) and the newly proposed models (A2Mid2E, A2Mid2E-Joint). A positive cost indicates a loss in performance.}
	\label{tbl:experiments}
\end{table*}

The audio clips are preprocessed as described in Section~\ref{sec:models} to obtain the input spectrograms. During training, one random 10-second snippet from each spectrogram is taken as input. We optimize the mean squared error, and use Pearson's correlation coefficient as the evaluation metric for emotion rating prediction. Each of the paths (A2E, A2Mid2E, A2Mid2E-Joint) is run ten times and the average correlation values are reported. Each run has a different seed which reshuffles the train-test split.

\subsection{Verifying our Basic Architecture}
\label{sec:basic}

Before going any further, we first want to verify that our VGG-style network model performs on par with comparable methods on the basic component tasks of predicting mid-level features from audio (A2Mid) and emotions from (given) mid-level features (Mid2E).

 For the A2Mid scenario (Table \ref{tbl:midlevel}), we train in two ways: first on the entire Mid-level features dataset with 8\% test set selected as described in \cite{aljanaki2018}. We call this A2Mid+. This is the result that should be directly compared to column 1. Second, we train only on the songs from the Soundtracks dataset with 20\% test set, and call this A2Mid. The column `Joint' in Table \ref{tbl:midlevel} gives the mid-level predictions produced by our A2Mid2E-Joint model. As can be seen, our models are broadly comparable to the results reported in \cite{aljanaki2018}, with A2Mid+ and \textit{Joint} performing slightly better, on average.
 
 Regarding the prediction of emotions from given mid-level feature annotations (Table \ref{tbl:experiments}, first two rows), there is not much space for deviation, as the models used by Aljanaki \cite{aljanaki2018} and us are very simple.
 
 \begin{table}[h]
    \begin{tabular}{l|cccc}
        \hline 
        \makecell{Mid-level\\feature} & Aljanaki & A2Mid+ & A2Mid & Joint \\
        \hline \hline 
        Melodiousness & 0.70 & 0.70 & 0.69 & 0.72 \\
        Articulation & 0.76 & 0.83 & 0.84 & 0.79 \\
        R. Stability & 0.46 & 0.39 & 0.39 & 0.34 \\
        R. Complexity & 0.59 & 0.66 & 0.45 & 0.46 \\
        Dissonance & 0.74 & 0.74 & 0.73 & 0.74 \\
        Tonal Stability & 0.45 & 0.56 & 0.61 & 0.63 \\
        Minorness & 0.48 & 0.55 & 0.51 & 0.57 \\
        \hline 
    \end{tabular}
    \caption{Correlation values for mid-level features predictions using our models, compared with those reported by Aljanaki et al.\cite{aljanaki2018}.} 
    \label{tbl:midlevel}
\end{table}

\subsection{Quantifying the Cost of Explainability}
\label{sec:cost}

We now compare our three model architectures (A2E, A2Mid2E, A2Mid2E-Joint) on the full task of predicting emotion from audio, 
We train on the Soundtracks dataset as described above, with 10 runs with randomly selected train-test 80:20 splits.
The A2E model serves as reference for the subsequent models with explainable linear layers. The results can be found in Table \ref{tbl:experiments}.

In the case of direct emotion prediction (A2E), the final layer is connected to 256 input nodes. However, in the A2Mid2E scheme, due to the fact that we introduce a bottleneck (viz. the 7 mid-level predictions) as inputs to the subsequent linear layer predicting emotions, our hypothesis is that doing so should result in a decrease in the performance of emotion prediction. We calculate this cost as the difference in correlation coefficients between the two models for each emotion. The results (rows A2E and A2Mid2E in Table \ref{tbl:experiments}) reflect the expected trend, but as can be seen, the decrease in performance is quite small (less than 7\% of the original correlation coefficient on average). 
\subsection{Joint Learning of Mid-level and Emotions}
\label{sec:joint}

Further improvements in the performance of the mid-level-based network can be obtained by training jointly on the mid-level and emotion annotations (as described in Section \ref{sec:a2ejoint}). This model reduces the cost even further, as can be seen in Table \ref{tbl:experiments}, row A2Mid2E-Joint. The decrease in performance is now less than 1.5\% of the correlation coefficients for the A2E case on average. We believe this is acceptable in view of the possibility of obtaining explanations from this network (see below).

\section{Obtaining Explanations}
\label{sec:iml}

Since the mapping between mid-level features and emotions is linear in both proposed schemes (A2Mid2E, A2Mid2E-Joint), it is now straightforward to create human-understandable explanations. Linear models can be interpreted by analyzing their weights: increasing a numerical feature by one unit changes the prediction by its weight. A more meaningful analysis is to look at the \textit{effects}, which are the weights multiplied by the actual feature values~\cite{molnar2019}. 
An effects plot shows the distribution, over a set of examples, of the effects of each feature on each target.
Each dot in an effects plot can be seen as the amount this feature contributes (in combination with its weight) to the prediction, for a specific instance. Instances with effect values closer to 0 get a prediction closer to the intercept (bias term).
Figure \ref{fig:effects} shows the effects of the model A2Mid2E-Joint. 

First we will show how this can be used to provide model-level explanations and then we will explain a specific example at the song level.

\subsection{Model-level Explanation}
\label{sec:modellevel}

\begin{figure}[h]
    \centering
		\includegraphics[width=\columnwidth]{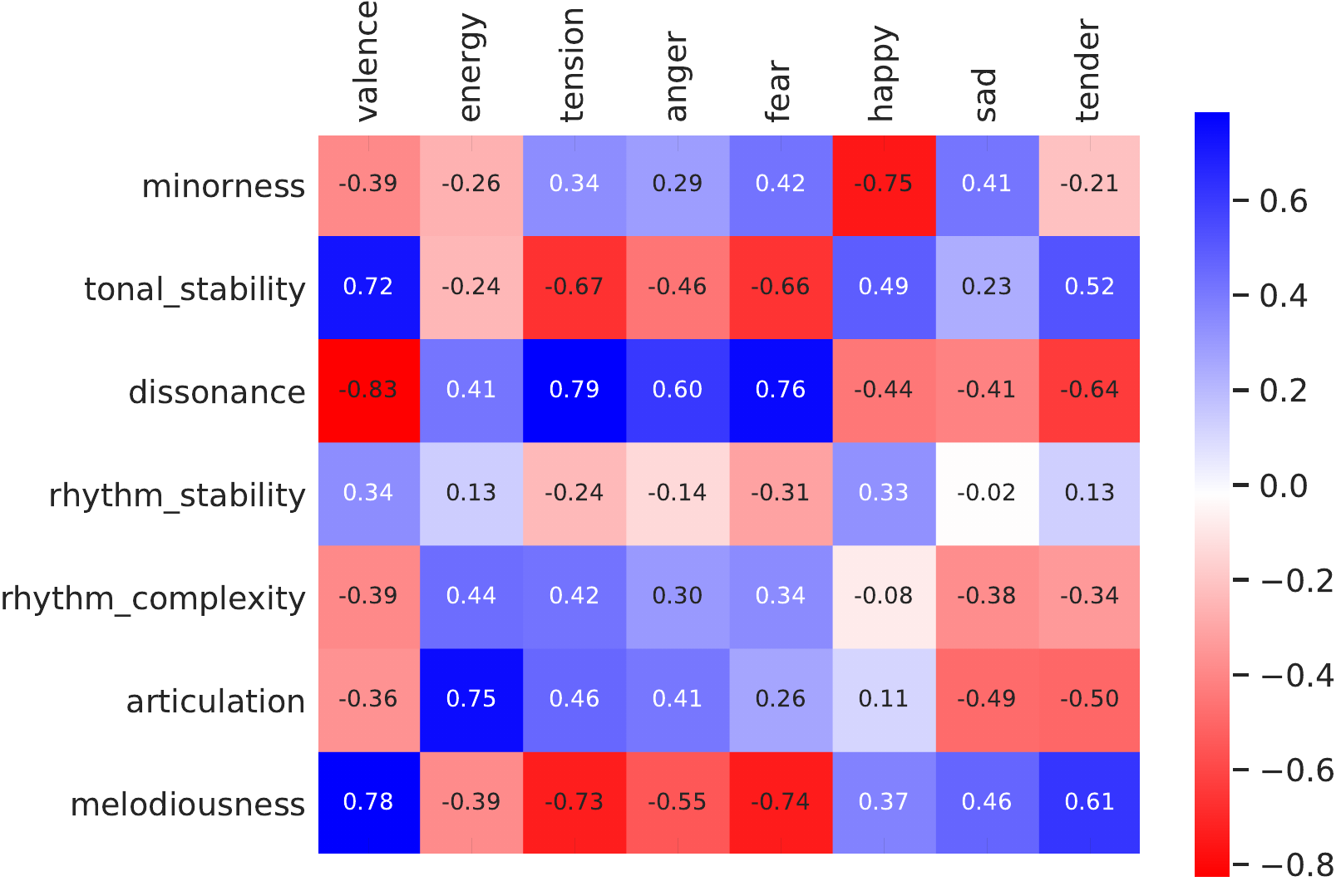}
		\caption{Pairwise correlation between mid-level and emotion annotations.}
		\label{fig:xcorr}
\end{figure}

\begin{figure}[h]
    \centering
		\includegraphics[width=\columnwidth, clip]{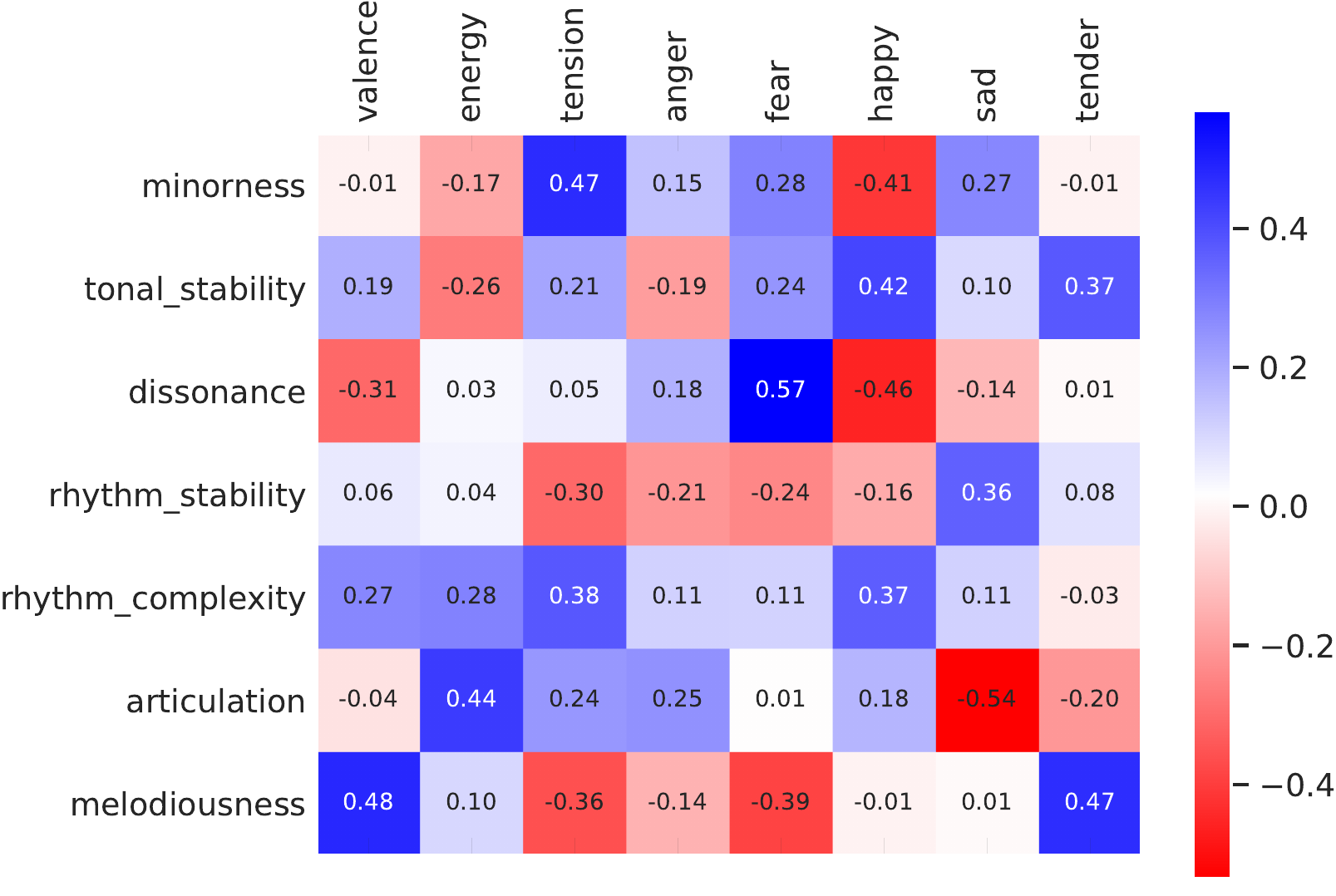}
		\caption{Weights from the linear layer of the `A2Mid2E-Joint' model.}
		\label{fig:joint_weights}
\end{figure}

Before a model is trained, the relationship between features and response variables can be analyzed using correlation analysis. 
The pairwise correlations between mid-level and emotion annotations in our data are shown in Figure~\ref{fig:xcorr}. When we compare this to the effect plots in Figure~\ref{fig:effects}, or the actual weights learned for the final linear layer (Figure~\ref{fig:joint_weights}) it can be seen that for some combinations (e.g., valence and melodiousness, happy and minorness) positive correlations go along with positive effect values and negative correlations with negative effect values, respectively. This is not a general rule, however, and there are several examples (e.g., tension and dissonance, energy and melody) where it is the other way around. The explanation for this is simple: correlations only consider one feature in isolation, while
learned feature weights (and thus effects) also depend on the other features and must hence be interpreted in the overall context. Therefore it is not sufficient to look at the data in order to understand what a model has learned.

To get a better understanding, we will look at each emotion separately, using the effects plot given in Figure~\ref{fig:effects}. In addition to the direction of the effect -- which we can also read from the learned weights in Figure~\ref{fig:joint_weights} (but only because all of our features are positive) -- we can also see the spread of the effect which tells us more about the actual contribution the feature can have on the prediction, or how different combinations of features may produce a certain prediction.

\subsection{Song-level Explanations}
\label{sec:songlevel}

\begin{figure*}[t]
		\centering
		\includegraphics[width=\textwidth]{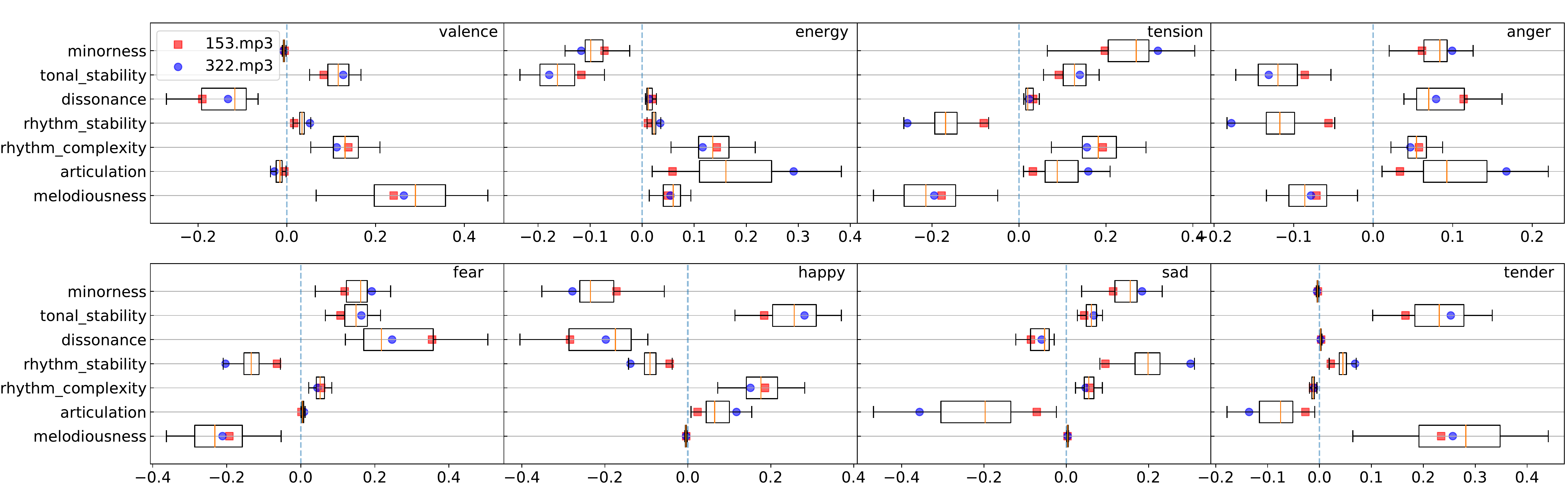}
		\caption{Effects of each mid-level feature on prediction of emotion. The boxplots show the distribution of feature effects of the model `A2Mid2E-Joint' helping us to understand the model globally. Additionally, two example songs (blue dots, red squares) are shown to provide song-level explanations (see Section \ref{sec:songlevel} for a discussion).}
		\label{fig:effects}
\end{figure*}

Effect plots also permit us to create simple example-based explanations that can be understood by a human. 
The feature effects of single examples can be highlighted in the effects plot in order to analyze them in more detail, and in the context of all the other predictions. To show an interesting case we picked two songs with similar emotional but different midlevel profiles. To do so we computed the pairwise euclidean distances between all songs in emotion ($d_{\text{E}}$) and midlevel space ($d_{\text{Mid}}$) separately, scaled both to the range $[0, 1]$ and combined them as $d_{\text{comb}} = d_{\text{E}} - (1 - d_{\text{Mid}})$. We then selected the two songs from the Soundtracks dataset that maximised $d_{\text{comb}}$. The samples are shown in Figure~\ref{fig:effects} as a red square (song \#153) and a blue dot (song \#322). The reader can listen to the songs/snippets by downloading them from the Soundtracks dataset page\footnote{\label{note_st}{https://www.jyu.fi/hytk/fi/laitokset/mutku/en/research/projects2/past-projects/coe/materials/emotion/soundtracks}}.

\begin{table}
    \centering
	\begin{tabular}{l|rr|rr}
	\hline 
	     & \multicolumn{2}{c|}{predicted} & \multicolumn{2}{c} {annotated} \\
      \hline
               & \#153  & \#322 & \#153  & \#322 \\
      \hline \hline 
      valence  & 0.28 & 0.39 & 0.38 & 0.46 \\
      energy   & 0.37 & 0.50 & 0.37 & 0.54 \\
      tension  & 0.40 & 0.46 & 0.50 & 0.56 \\
      anger    & 0.28 & 0.23 & 0.15 & 0.22 \\
      fear     & 0.41 & 0.27 & 0.18 & 0.28 \\
      happy    & 0.17 & 0.21 & 0.17 & 0.17 \\
      sad      & 0.20 & 0.23 & 0.27 & 0.28 \\
      tender   & 0.18 & 0.23 & 0.10 & 0.10 \\
      \hline 
	\end{tabular}
	\caption{Emotion prediction profiles for the two example songs \#153 and \#322.}
	\label{tab:emotionpredictions}
\end{table}

As can be seen from Figure~\ref{fig:effects} and from the emotion prediction profile of the two songs (see Table \ref{tab:emotionpredictions}), both songs have relatively high predicted values for \textit{tension} and \textit{energy},
but apparently for different reasons: song \#322 more strongly relies on ``minorness" and ``articulation" for achieving its ``tense" character; on the other hand, its rhythmic stability counteracts this more strongly than in the case of song \#153. The higher score on the ``energy" emotion scale for \#322 seems to be primarily due to its much more articulated character (which can clearly be heard: 153 is a saxophone playing a chromatic, harmonically complex line, 322 is an orchestra playing a strict, \textit{staccato} passage).



\section{Discussion and Conclusion}
\label{sec:discussion}

Model interpretability and the possibility to obtain explanations for a given prediction are not ends in themselves.
There are many scenarios where one may need to understand why a piece of music was recommended or placed in a certain category. Concise explanations in terms of mid-level features would be attractive, for example, in recommender systems or search engines for `program music' for professional media producers, where mid-level qualities could also be used as additional search or preference filters\footnote{A demonstration can be found in \url{https://shreyanc.github.io/ismir_example.html}}. As another example, think of scenarios where we want a music playlist generator to produce a music program with a certain prevalent mood, but still maintain musical variety within these limits. This could be achieved by using the mid-level features underlying the mood/emotion classifications to enforce a certain variability, by sorting or selecting the songs accordingly.

There are several obvious next steps that need to be taken in the research. The first is to extend this analysis to a larger set of diverse datasets and emotion-related dimensions.\footnote{In fact, we do have preliminary results on a second dataset -- the  \emph{MIREX-like Mood Dataset)} of ~\cite{panda2013}, which is also covered by the Mid-level Perceptual Features Dataset of \cite{aljanaki2018} and differs from \textit{Soundtracks} in that is comes with discrete mood labels. The results confirm the general trends reported in the present paper, but because of the different emotion/mood encoding scheme, further optimisations on our models may still improve the results further.}

Second, we plan to extend the models and sets of perceptual features.
One rather obvious (and obviously relevant) perceptual dimension that is conspicuously missing from our (Aljanaki \& Soleymani's) set of mid-level features is \textit{perceived speed} (which is not the same as tempo). Adding this intuitive musical dimension is an obvious next step towards improving our model. Of course, this will require an appropriate ground truth for training.

Generally, the relation between the space of musical qualities (such as our mid-level features) and the space of musically communicated emotions and affects deserves more detailed study. A deeper understanding of this might even give us means to control or modify emotional qualities in music by manipulating mid-level musical properties.

%

\section{Acknowledgments}

This research has received funding from the European Research Council (ERC) under the
European Union’s Horizon 2020 research and innovation programme under grant agreement
No.~670035 (project ``Con Espressione'').

\bibliography{ISMIRtemplate}

%
%
%
%

\end{document}